\documentclass{aastex62}

\graphicspath{{./}{figures/}}
\usepackage{epstopdf}

\begin{document}

\title{Solar Image Restoration with the Cycle-GAN Based on Multi-Fractal Properties of Texture Features}

%% Note that the corresponding author command and emails has to come
%% before everything else. Also place all the emails in the \email
%% command instead of using multiple \email calls.
\author[0000-0001-6623-0931]{Peng Jia}
\email{robinmartin20@gmail.com}
\affil{College of Physics and Optoelectronics,  Taiyuan University of Technology, Taiyuan, 030024, China}
\affiliation{Department of Physics, Durham University, South Road, Durham, DH1 3LE, UK}
\affiliation{Key Laboratory of Advanced Transducers and Intelligent Control Systems, Ministry of Education and Shanxi Province, Taiyuan University of Technology, Taiyuan, 030024, China}

\author{Yi Huang}
\affiliation{College of Physics and Optoelectronics,  Taiyuan University of Technology, Taiyuan, 030024, China}

\author{Bojun Cai}
\affiliation{College of Physics and Optoelectronics,  Taiyuan University of Technology, Taiyuan, 030024, China}

\author{Dongmei Cai}
\affiliation{College of Physics and Optoelectronics,  Taiyuan University of Technology, Taiyuan, 030024, China}

%% Note that RNAAS manuscripts DO NOT have abstracts.
%% See the online documentation for the full list of available subject
%% keywords and the rules for their use.
\begin{abstract}
Texture is one of the most obvious characteristics in solar images and it is normally described by texture features. Because textures from solar images of the same wavelength are similar, we assume texture features of solar images are multi-fractals. Based on this assumption, we propose a pure data-based image restoration method: with several high resolution solar images as references, we use the Cycle-Consistent Adversarial Network to restore burred images of the same steady physical process, in the same wavelength obtained by the same telescope. We test our method with simulated and real observation data and find that our method can improve the spatial resolution of solar images, without loss of any frames. Because our method does not need paired training set or additional instruments, it can be used as a post-processing method for solar images obtained by either seeing limited telescopes or telescopes with ground layer adaptive optic system.
\end{abstract}

%% Start the main body of the article. If no sections in the
%% research note leave the \section call blank to make the title.
\section{Introduction} \label{sec:intro}
The imaging process of optical telescopes can be modeled by equation \ref{eq:equation1}:\\
\begin{equation} \label{eq:equation1}
Img(x,y)=Obj(x,y)\ast PSF(x,y)+Noise(x,y),
\end{equation}
where $Obj(x,y)$ and $Img(x,y)$ are the original and observed images, $\ast$ is the convolutional operator, $PSF(x,y)$ is the point spread function (PSF) of the whole optical system and $Noise(x,y)$ stands for the noise from the background and the detector. During real observations, many different effects will introduce variable $PSF(x,y)$ and $Noise(x,y)$. These effects will make the $Img(x,y)$ different from the $Obj(x,y)$ and strangle further scientific researches.\\

For ground based solar observations, because the exposure time is short (dozens of millisecond) and the field of view is big (comparing to the isoplanatic angle), the atmospheric turbulence, thermal and gravity deformations of the optical system will introduce $PSF(x,y)$ with highly spatial and temporal variations. Even with the help of adaptive optics and active optics systems, the residual error will still introduce a variable $PSF(x,y)$. The variable PSF of solar images is different from that of ordinary night time astronomy observations and it is often called short exposure PSF, because the exposure time is only dozens of milliseconds. The short exposure PSF can not be described by any contemporary analytical PSF models such as the Moffat or the Gaussian model and it is the main limitation for ground based solar observations.\\

Several different image restoration methods have been proposed to reduce the effects brought by the short exposure PSF and increase the spatial resolution of astronomical images, such as the blind deconvolution algorithm \citep{Jefferies1993}, the speckle reconstruction algorithm \citep{Labeyrie1970, Luehe1993}, the phase diversity algorithm \citep{Paxman1992,Paxman1996,Matt1994} and the multi-object multi-frame blind deconvolution algorithm \citep{Noort2005}. These methods have different hypothesis or prior value of the $PSF(x,y)$ or the $Img(x,y)$, including wavefront measurements or assuming the image is invariant between different frames, and have achieved remarkable performance.\\

Texture is fundamental characteristic of an image and it describes the grey scale spatial arrangement of images. Normally texture features are used to evaluate textures. In our recent paper \citep{Huang2019}, we have shown the multi-fractal properties of texture features in solar images of different wavelengths. Based on the results from that paper, we use a Cycle-Consistent Adversarial Network (CycleGAN) to restore solar images with the multi-fractal properties as regularized condition in this paper. Our method can restore arbitrary number of  solar images obtained by the same telescope within a few days with only several high resolution images as references. We will discuss the multi-fractal property of texture features in Section 2 and introduce our method in Section 3. In Section 4, we will show the performance of our method with real and simulated observation data. We will make our conclusions and anticipate our future work in Section 5.\\

\section{The Multi-Fractal Property of Texture Features In Solar Images}
\label{sec:solarfractal}
\begin{figure}
\begin{center}
\includegraphics[scale=1.0,angle=0]{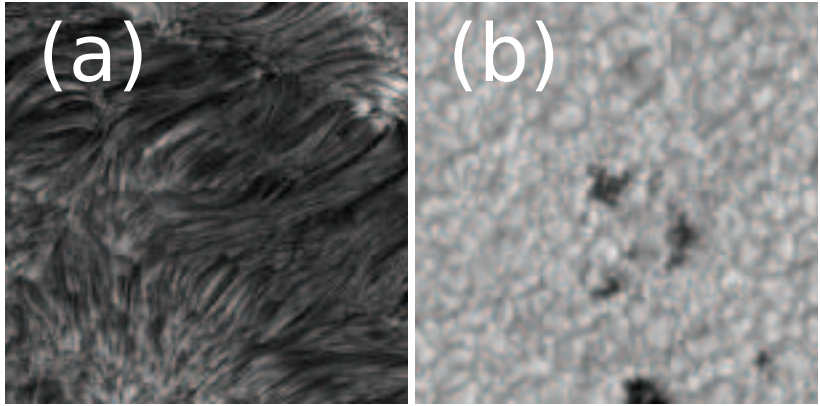}
\caption{Examples of high resolution solar images obtained by (a) H-alpha and (b) TiO filters of the New Vacuum Solar Telescope (NVST). Besides the big structures, we can notice that there are a lot of textures in these images. People can easily tell the difference between images from these two wavelengths, because textures in the same wavelength are relatively stable since high resolution observations are carried out by human being.\label{image}}
\end{center}
\end{figure}

Textures are mostly related to spatially repetitive structures which are formed by several repeating elements \citep{castelli2002image}.  Similar to other natural images, solar images also have a lot of textures, such as the granulation in TiO and the filament in H-alpha, as shown in Figure~\ref{image}.  The texture feature is a description of the spatial arrangement of the gray scale in an image and it is usually used to describe the regularity or coarseness of an image. Manually designed texture features have been successfully used to describe the arrangement of texture constituents in a quantitative way \citep{tamura1978textural,manjunath1996texture,li2019automatic}. However, textures in solar images are not arranged in a regular or periodic way, which makes it hard to design adequate texture features by hand. According to our experience, textures in solar images have the following properties:\\
1. In the same wavelength, textures from different solar images are similar.\\
2. For the same solar image, the shape variation of textures in the same spatial scale satisfies the same statistical law. For example, filaments are bending with a smooth curve instead of a polyline.\\

These properties indicate us that although texture features are not organized in a regular way for solar images, the relative weights of different texture features are stable, which means if we measure the relative weights of texture features in a statistical way, the probability distribution in the same wavelength should be the same. We can use multi-fractal properties to describe texture features in solar images \citep{Jia2014, peng2017discrimination}. The multi-fractal property of texture features means the spatial distribution of textures in solar images (coded by texture features) satisfies the same continuous power spectrum and for different scales, the exponents of the spectrum are different.\\

Because textures are just appearances of the physical process behind and the physical process that generates the textures does not change, the multi-fractal property of texture features are valid for all solar images obtained in the same wavelength, i.e., texture features of solar images in the same wavelength satisfy the same power spectrum. Recent works suggest that neural network have very good property in representing complex functions.\\ 
In this paper, we further try to take advantage of this property and propose to use neural networks to evaluate multi-fractal properties from solar images. The multi-fractal properties are encoded in the neural network and can be visualized by feature maps of each layer. In our recent papers \citep{Huang2019}, we show the multi-fractal properties of G-band images. In this paper, the multi-fractal properties are used as regularized condition of the CycleGAN, just as other regularization conditions used in traditional deconvolution algorithm, such as: the total-variation condition in deconvolution algorithms. So we do not try to extract the multi-fractal property of texture features directly, instead we use many high resolution images to represent it in a statistical way.\\ 

In real observations, we can obtain a lot of high resolution solar images through speckle reconstruction, phase diversity, multi-object multi-frame deconvolution or observations with diffraction-limited adaptive optic systems, such as the single-conjugate or multi-conjugate adaptive optics systems \citep{Rao2018}. The spatial resolution of these images are around the diffraction limit of the telescope and can reveal the highest spatial frequency that the observation data have. As the texture features from high resolution solar images are just realization of the theoretical multi-fractal property, we can use multi-fractal properties of texture features from high resolution images as a restriction condition for image post-processing methods, as we will discuss in the next section.\\

\section{The CycleGAN for Image Restoration with the Multi-fractal Property} \label{sec:ccgan}
\subsection{Introduction of the CycleGAN}\label{sec:intrcycle}
The deep convolutional neural network (DCNN) is a type of deep learning framework and is widely used in image restoration \citep{xu2014deep,wieschollek2016end-to-end,zhang2017learning}. For solar images, several different DCNNs have been proposed for image restoration or enhancing \citep{Diazbaso2018, Asensio2018}. These methods are based on supervised learning, which requires pairs of high resolution images and blurred images as training set to model the degradation process, i.e. the $PSF(x,y)$. However, for real observations, obtaining the training set is hard. Besides, the number and diversity of images in the training set are usually not large enough to represent different image degradation processes. A trained DCNN will output unacceptable results, when blurred images have a different $PSF(x,y)$ than that of the training set. The requirement of many paired images in the training set limits wider application of these image restoration methods.\\

The generative adversarial network (GAN) is a generative model \citep{goodfellow2014generative}, which contains two DNNs: a generator {\bfseries G} and a discriminator {\bfseries D}. Given real data set {\bfseries R}, {\bfseries G} tries to create fake data that looks like the genuine data from {\bfseries R}, while {\bfseries D} tries to discriminate the fake data and the genuine data. The GAN can be trained with back-propagation algorithm effectively, when there are only limited training data. For image restoration of galaxies, the GAN has been successfully trained with only 4105 pairs of training images \citep{Schawinski2017}. However, as we discussed above, the GAN also models the degradation process from these training images. For solar observations, because the atmospheric turbulence induced short exposure PSF is much more complex than the long exposure PSF, the number of training images required in the GAN will be greatly increased and the performance of GAN will be strongly influenced by limited training data.\\

Limited training data is also a problem in other image related tasks. \citet{zhu2017unpaired} propose the Cycle-Consistent Adversarial Network (CycleGAN) to solve this problem. The CycleGAN is an unsupervised learning algorithm, which contains a pair of Generative Adversarial Networks (GAN). Given two sets of images, one GAN learns the image mapping and the other GAN learns the inverse mapping. Under the constrain condition that the mapped image after inverse mapping should be similar to itself and vise versa (cycle consistency loss), the CycleGAN can restore blurred images directly with high resolution images as references. It should be noted that, when used for image restoration, although the supervised DCNN, the GAN and the CycleGAN all try to learn the restoration function, the CycleGAN has very different hypothesis than that of other methods. The CycleGAN is constrained by the probability distribution of data (multi-fractal property of texture features in this paper), while other methods are constrained by the blur properties contained in pairs of training images. The detail structure of the CycleGAN used in this paper will be discussed in the next subsection.\\

\subsection{Structure of the CycleGAN for Solar Image Restoration} \label{sec:cyclegan}
\begin{figure}
\begin{center}
\includegraphics[scale=0.5,angle=0]{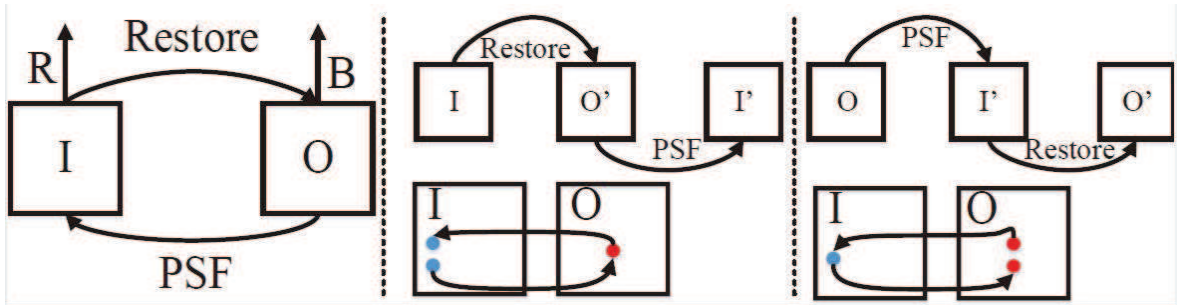}
\caption{The architecture of the CycleGAN used in this paper. The left figure shows the basic structure of CycleGAN. I stands for the observation image and O stands for the reference high resolution image. Restore and PSF stand for the two functions to be learned: the restore function and the PSF. $D_I$ and $D_O$ are two discriminators which are used to evaluate the generator output. The middle and right figure stand for the two learning processes in the CycleGAN. The blue dots are blurred images and the red dots are high resolution images. The CycleGAN will restore I to O' and then will blur O' to I' and vise versa from O to O'. The cycle-consistency loss is introduced to make sure that the above translation will not change the image. 
\label{fig:cyclegan}}
\end{center}
\end{figure}
 \begin{figure}
\gridline{\fig{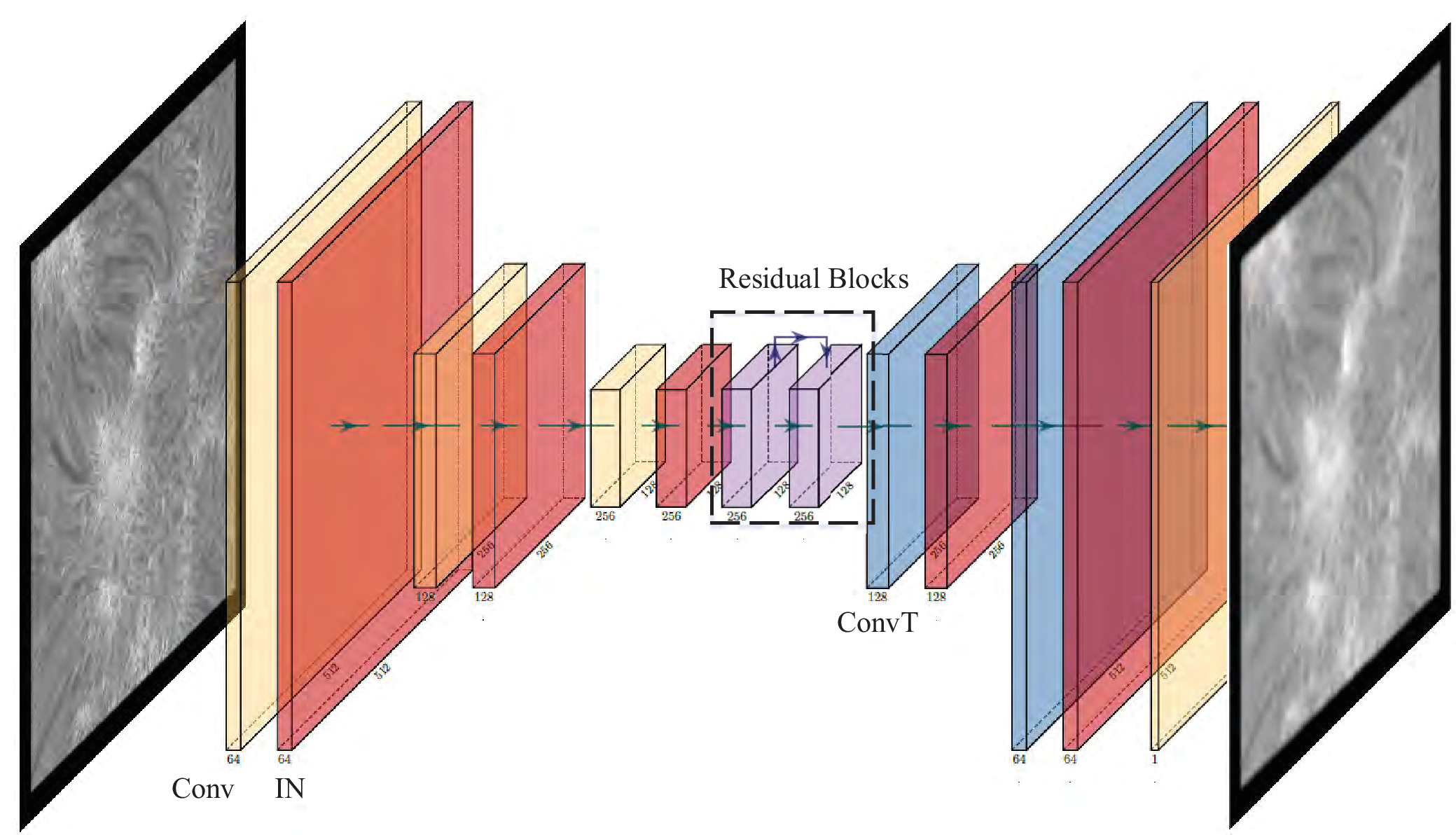}{0.6\textwidth}{(a)}
          \fig{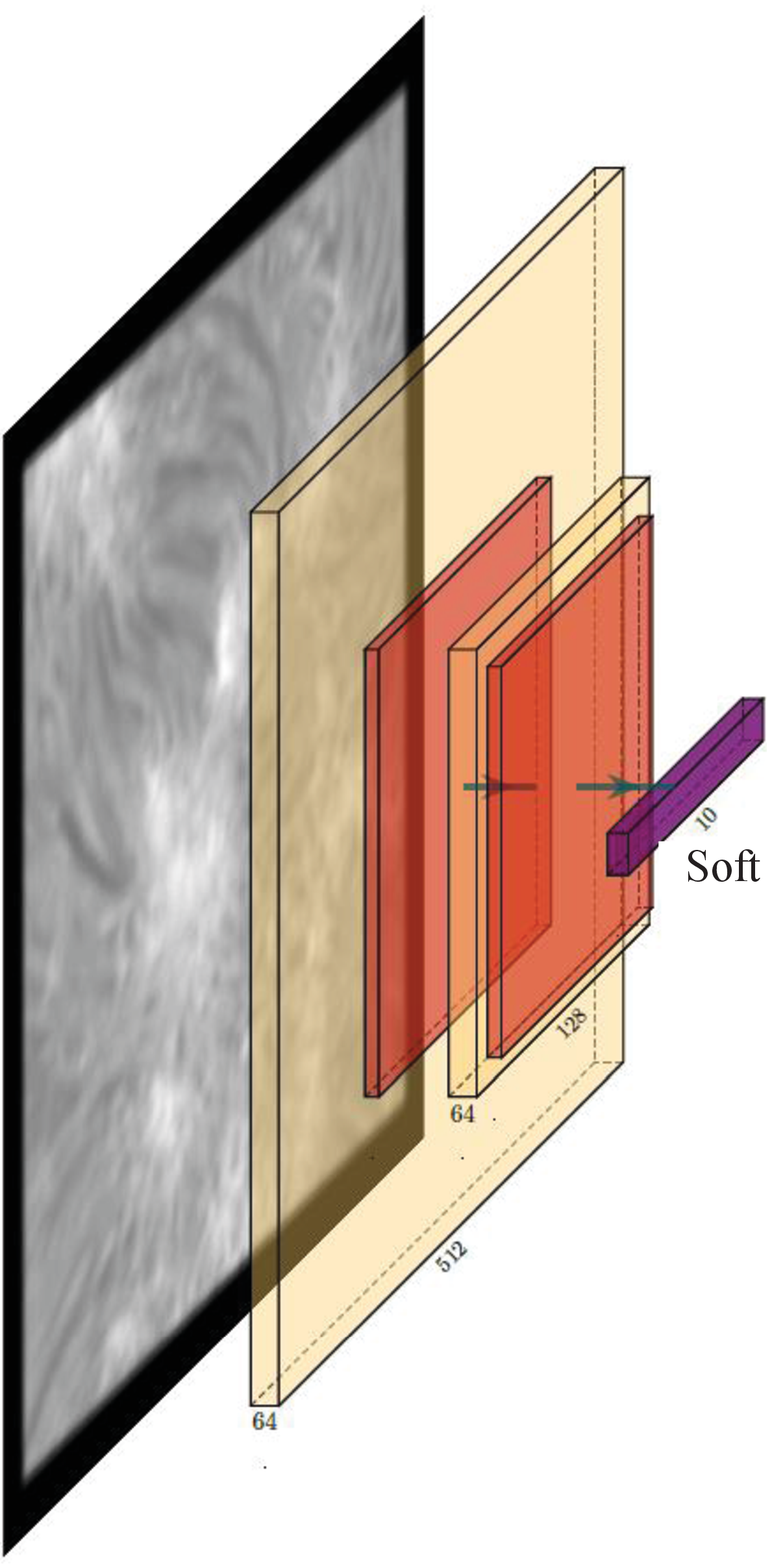}{0.17\textwidth}{(b)}}
\caption{The architecture of the generator in figure (a) and the discriminator in figure (b). The generator is shown in figure (a), which consists of convolutional layers (Conv in yellow color), instance normalization layers (IN in red color), Conv-transpose layers (convT in blue color) and residual blocks (residual blocks in purple color). The discriminator is shown in figure (b), which consists of convolutional layers, instance normalization layers and softmax layers (in dark purple). The high resolution image $O$ will be transformed to the blurred image $I$ through the generator and the blurred image will be sent to the discriminator $D_I$ for discrimination.  There are two of sets of the generator and the discriminator in the CycleGAN and they have the same structure.\label{NN}}
\end{figure}
The structure of the CycleGAN used in this paper is shown in Figure~\ref{fig:cyclegan}\footnote{The complete code used in this paper is written in Python programming language (Python Software Foundation) with the package Pytorch and can be downloaded from \href{JPwebsite}{http://aojp.lamost.org}.}. Because short exposure PSFs for solar observation have very complex structure, we use a very deep CNN as generator as shown in Figure~\ref{NN} and Appendix~\ref{app}. This generator is inspired by \citet{zhu2017unpaired}. However, since the CycleGAN will learn the short exposure PSF which has complex structures, we modify it and use smaller convolution kernels here to increase its representation ability. For the discriminator, we use an ordinary CNN that is normally used in image style transfer \citep{isola2017image-to-image,Yi2017DualGAN,zhu2017unpaired}.\\

As shown in Figure~\ref{fig:cyclegan}, $I$ and $O$ are blurred images and high resolution images. One GAN in the CycleGAN tries to learn the restoration function and it has a generator $Restore$ and a discriminator $D_O$, which can be written as: $Restore: I \rightarrow O$. The other GAN tries to learn the PSF and it has a generator $PSF$ and a discriminator $D_I$, which can be written as $PSF: O \rightarrow I$. We apply ordinary adversarial loss to both of these two GANs as:\\
\begin{eqnarray} \label{eq:GANloss}
 \mathcal{L}_{GAN}(Restore, D_{O}, I, O) = \mathbb{E}_{O\sim p_{data}(O)}[log D_{O}(O)]+\mathbb{E}_{I\sim p_{data}(I)}[log(1- D_{O}(Restore(I)))],\\
 \mathcal{L}_{GAN}(PSF,D_{I},I,O) =\mathbb{E}_{I\sim p_{data}(I)}[log D_{I}(I)]+\mathbb{E}_{O\sim p_{data}(O)}[log(1- D_{I}(PSF(O)))] ,
\end{eqnarray}
where $\mathbb{E}_{I\sim p_{data}(I)}$ stands for the expectation in the probability space of $data(I)$ and vise versa. These two adversarial loss can make the distribution generated by the above two generators close to the real distribution of $I$ or $O$. However because the generator is very complex, with the adversarial loss along, the generator would learn other mappings that also match the distribution of $I$ or $O$. To further restrict the space of possible mapping functions, we use the cycle consistency loss $\mathcal{L}_{cyc}(Restore,PSF)$ to constrain the solution space:\\
\begin{eqnarray} \label{eq:cycloss}
\mathcal{L}_{cyc}(Restore,PSF)=\mathbb{E}_{O\sim p_{data}(O)}[\Vert Restore(PSF(O))-O\Vert_{1}]+\mathbb{E}_{I\sim p_{data}(I)}[\Vert PSF(Restore(I))-I\Vert_{1}],
\end{eqnarray}
where $\Vert\Vert_{1}$ stands for the 1-norm. The cycle consistency loss guarantees that, for each image, the CycleGAN will bring it back to the original value:\\
\begin{eqnarray} \label{eq:cyclossfor}
O \rightarrow PSF(O)  \rightarrow Restore(PSF(O))  \approx  O\\
I \rightarrow Restore(I)  \rightarrow PSF(Restore(I) )  \approx  I
\end{eqnarray}

Because the image restoration algorithm should not flip the gray value between different pixels, we use the identity loss to constrain the contrast of the image, preventing from rapid change of gray scale between different pixels.\\
\begin{eqnarray} \label{eq:idenity}
\mathcal{L}_{identity}(Restore,PSF)=\mathbb{E}_{O\sim p_{data}(O)}[\Vert PSF(O)-O\Vert_{1}]+\mathbb{E}_{I\sim p_{data}(I)}[\Vert Restore(I)-I\Vert_{1}]
\end{eqnarray}

At last, we calculate the total variation of $O$ and $I$ and use them as the total variation loss to improve the image quality and reduce the artifacts generated by the CycleGAN, \\
\begin{eqnarray} \label{eq:tvloss}
\mathcal{L}_{TV}(Restore,PSF)=\mathbb{E}_{I\sim p_{data}(I)}[\Vert\nabla_{h}Restore(I)\Vert_{2}+\Vert\nabla_{w}Restore(I)\Vert_{2})]\nonumber \\
+\mathbb{E}_{O\sim p_{data}(O)}[\Vert\nabla_{h}PSF(O)\Vert_{2}+\Vert\nabla_{w}PSF(O)\Vert_{2})],\end{eqnarray}
where $\Vert\Vert_{2}$ stands for the 2-norm,  $\nabla_{h}$ and $\nabla_{w}$ are horizontal and vertical gradient of these images. We calculate the weighted summation of the above loss functions and use the function defined below to train the CycleGAN, where $\lambda$ is the relative weight of the cycle consistent loss.\\
\begin{eqnarray} \label{eq:FullLoss}
\mathcal{L}(Restore,PSF)= \mathcal{L}_{GAN}(Restore,D_{O},I,O) + \mathcal{L}_{GAN}(PSF,D_{I},I,O) \nonumber \\
+\lambda \mathcal{L}_{cyc}(Restore,PSF) + \mathcal{L}_{identity}(Restore,PSF) + \mathcal{L}_{TV}(Restore,PSF).
\end{eqnarray}

\subsection{Other Restriction Conditions for the CycleGAN in Image Restoration} \label{sec:cyclegan}
Because the CycleGAN tries to model the degradation process according to the statistical probability distribution of texture features in images, the restriction of this model should lie both in the image and the degradation process. First of all, as the CycleGAN does not have any paired training images as supervisions and it is supervised in the form of $O$ and $I$, where $O$ and $I$ are high resolution images and blurred images respectively, they should satisfy the following properties. For $O$:\\
1. High resolution images need to have the same or smaller pixel scale than that of the blurred images. Then we will down sample all the high resolution images to images of the same pixel scale as that of blurred images in data set $O$.\\
2. The apparent structures should be removed from high resolution images, albeit keeping it small enough. Because the apparent structures, such as sunspots have different textures and will change the multi-fractal properties.\\
3. We need enough images with textures to represent the multi-fractals in a statistical way. According to our experience, at least 100 frames of reference images with $256 \times 256$ pixels are required, however it is much smaller than ordinary DNN based image restoration method.\\

As the CycleGAN is to model the restoration function and the PSF, which have very strong spatial and temporal variations, we need to set several restrictions in $I$ to make the CycleGAN robust in real applications:\\
1. When restoring a single frame of solar image, it would be better to divide it into smaller images with size of around dozens of arcsec. \\
2. For several continuous frame of solar images, it would be better to cut the interested areas (with size of around dozens of arcsec) from these images and directly restore these temporal continuous small images with the CycleGAN.\\
3. To reduce image processing time, it would be better to use texture-rich small images that are cut from blurred images to train the CycleGAN. After training, the $Restore$ can be directly used to restore all the blurred images.\\

Last but not least, reference images and blurred images need to have the same multi-fractal property, so the images taken in the same band of wavelength within a few days is strict enough for our method. We tested our method with real observation data and found that the neural network used in this paper is complex enough for image restoration task, because we did not find any images that the CycleGAN failed to restore. With the above restrictions, the CycleGAN can be effectively trained through several thousand iterations. In the next Section, we will show implementations of our algorithm.\\
\section{Implementations of the CycleGAN for Image Restoration} \label{sec:imp}
\subsection{Performance Evaluation with Simulated Observation Data}
Over-fitting is a major problem that would limit the performance of our algorithm. Because the CycleGAN is a generative model, overfitting will make $Restore$ remember the structure of high resolution images and generate fake structures during image restorations. To test our algorithm, we use the CycleGAN to restore simulated blurred images. There are two sets of images used in this paper: observations carried out in H-alpha wavelength between 02 April 2018 and 03 April 2018 and observations carried out in TiO band on 17 November 2014. These two sets of data are both observed by the NVST  \citep{liu2014new}:  the H-alpha data was observed in 656.28 nm and bandpass of 0.025 nm with pixelscale of 0.136 arcsec and the TiO data was observed in 705.8 nm and bandpass of 1 nm with with pixelscale of  0.052 arcsec \footnote{For more details, please refer to: http://fso.ynao.ac.cn/cn/introduction.aspx?id=8}. These images are restored by speckle reconstruction methods \citep{Li2015High}.\\
From 5 high resolution H-alpha wavelength images, we crop 500 images with size of $256\times 256$ pixels as references and crop another 5 images with size of $256\times 256$ as test images. From 5 high resolution TiO solar images, we crop 500 images with size of $256\times 256$ pixels as references and crop another 5 images with size of $256\times 256$ as test images. According to real observation conditions, we use Monte Carlo method to simulate several high fidelity atmospheric turbulence phase screens with $D/r_0$ of 10 for H-alpha wavelength and 4 for TiO data \citep{jia2015simulation,jia2015real-time}, where $D$ stands for the diameter of telescope and $r_0$ stands for the coherent length of atmospheric turbulence. We calculate simulated short exposure PSFs with these phase screens through far field propagation \citep{Basden2018The}. At last, we convolve test images and temporal-continuous PSFs to generate 100 simulated blurred images as simulated blurred H-alpha and TiO data.\\

\begin{figure}
\centering
\includegraphics[scale=1.0,angle=0]{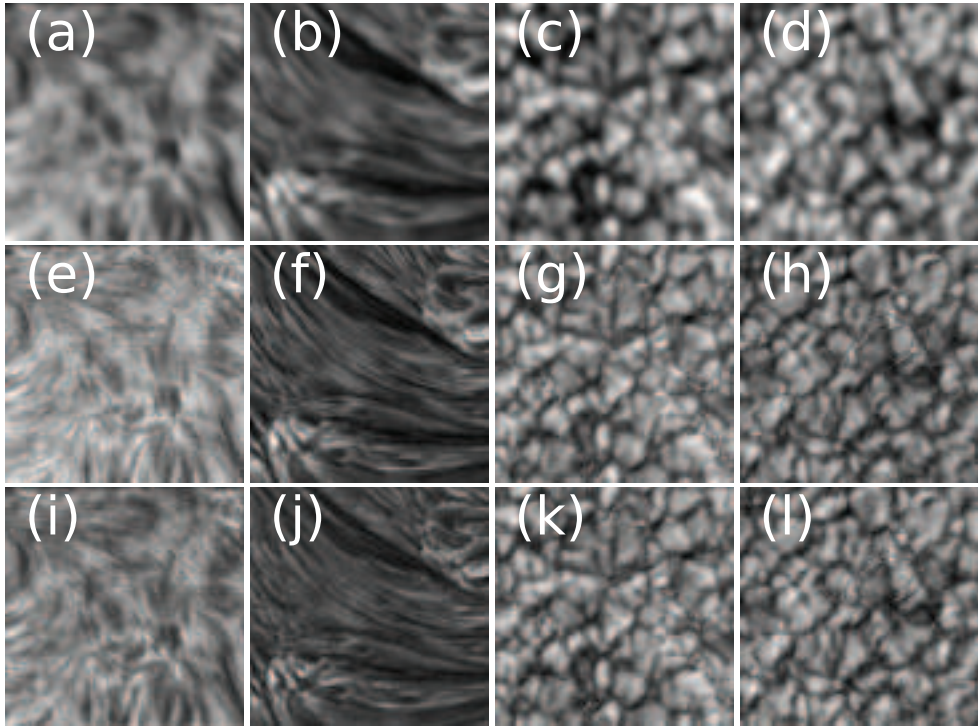}
\caption{Simulated short exposure images, restored images and the original high resolution images. The figures in the top row show the simulated blurred images, the figures in the middle row show the original high resolution images and images in the bottom row show the images restored by our method. The left two columns are images in H-alpha wavelength and the right two columns are images in TiO. From these figures, we can find the spatial resolution is increased by our method and comparing with the original high resolution images, there are no obvious artefacts in the restored images.\label{shortexposure}}
\end{figure}

These simulated blurred images and high resolution images in the same wavelength are used to train the CycleGAN with 2000 iterations. Because all the simulated blurred images have the original high resolution images, we can compare them to test our algorithm. The results are shown in the figure \ref{shortexposure}. We carefully check these images and find that the resolution of the restored images has been improved and there are no observable difference between the restored images and the the original images. We have also calculated the median filter-gradient similarity -- MFGS \citep{Deng2017} of the simulated blurred images, the original high resolution images and the restored images to further test our method. For H-alpha data, the MFGS is increased from $0.75\pm 0.05$ to $0.81 \pm 0.04$, while the mean MFGS of the original images is $0.81 \pm 0.04$. For TiO data, the mean MFGS is increased from $0.81 \pm 0.03$ to $0.82 \pm 0.02$, while the mean MFGS of the original images is $0.82 \pm 0.03$. According to these results, we can find that the image quality has been increased by our method. \\
\subsection{Performance Evaluation with Real Observation Data}
In this part, we use real observation data from NVST to evaluate performance of the CycleGAN and also show our recommendations of how to use the CycleGAN in real applications. According to the size of isoplanatic angle, an image should be cut into small images for restoration. However considering the processing speed, we use images with size of $256 \times 256$ pixels (equivalent to around $0.5\times0.5$ arcmin) for restoration. It is much larger than the isopanatic angle and the restoration results will drop down slightly, however it is a trade-off we have to make. In real applications, there are two scenes of image restoration: restoration of several continuous solar images in a small interested region or restoration of a single frame image with relative large size.\\

\begin{figure}
\begin{center}
\includegraphics[scale=1.0,angle=0]{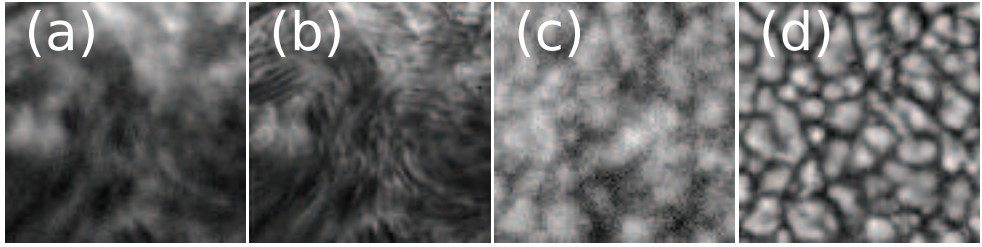}
\caption{Two frames of real observation images and their corresponding restored images. The left two images are from H-alpha wavelength and the right two images are from TiO. It is obvious that restored images have better quality. An animated versions of these figure are available in the on line version of this paper. \label{realData}}
\end{center}
\end{figure}

\begin{figure}
\begin{center}
\includegraphics[scale=0.5,angle=0]{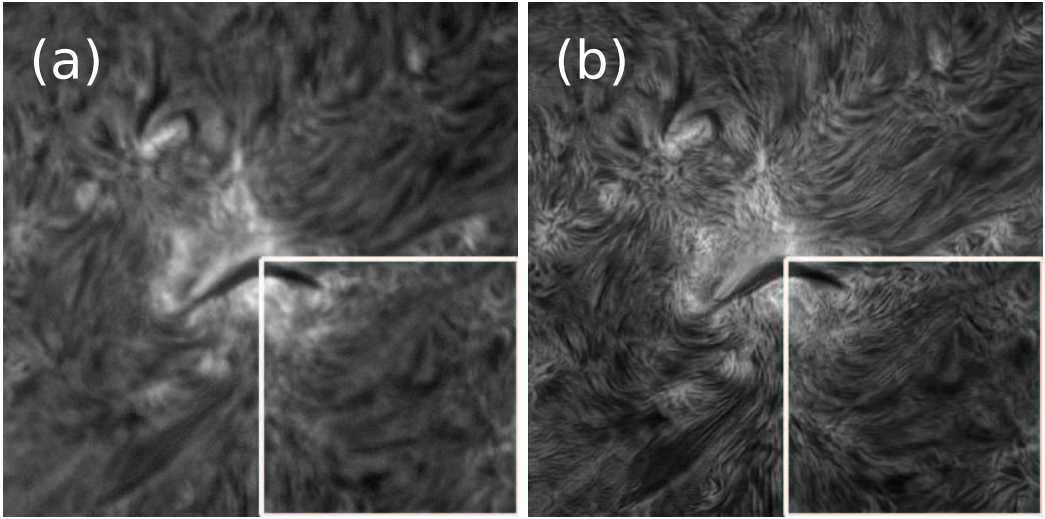}
\caption{Two frames of real observation images and their corresponding restored images. The size of these two images is $1024 \times 1024$ pixels. It is obvious that restored images have higher spatial resolution. The MFGS is increased from 0.78 to $0.89 \pm 0.003$ in the reference region and the MFGS in the test region of the restored image is $0.89$.\label{realData2}}
\end{center}
\end{figure}

For the first scene, we use both H-alpha images observed between 02 April and 03 April 2018 and TiO images observed on 19 August 2017 with small size to test our algorithm. For each wavelength, 500 frames of images with size $256\times 256$ pixels are extracted from 5 frames of blurred image with size of $1024\times 1024$ pixels as $I$ and 500 frames of images with size $256 \times 256$ pixels are extracted from one speckle reconstructed image with size of $1024\times 1024$ pixels as $O$. Because the CycleGAN is deep and complex, the maximal number and size of the reference image and the blurred image are actually limited by the computer\footnote{In this paper, we use a computer with two Nvidia GTX 1080 graphics cards, 128 GB memory and two Xeon E5 2650 processors. It will cost 4498 seconds to train the CycleGAN with 6000 iterations.}. After training, we can use the $Restore$ to directly restore two interested regions with size of $256\times 256$ pixels in the blurred images. It costs around 1 minutes to process 333 frames of these blurred images. Two frames of restored images are shown in Figure ~\ref{realData} and interested readers can also find animated versions of these figures in the online version of this paper. They are 100 frames of blurred solar images from H-alpha and 79 frames from TiO before and after restoration, alone with their MFGS values. From these figures, we can find that the resolution of the restored images have been improved. The mean MFGS is increased from $0.65 \pm 0.01$ to $0.89 \pm 0.004$ for TiO data and from $0.75 \pm 0.03$ to $0.89 \pm 0.007$ for H-alpha data.\\
For the second scene, we test the performance of our algorithm in big images. A blurred image from the NVST is shown in the panel of Figure~\ref{realData2}. The validation part is the image in the white box and the rest of this image is used as the training part. We extract 500 frames of images with size of $256\times 256$ from the training part as $I$ and use 500 frames of high resolution images used above as $O$ in the CycleGAN. After 6000 iterations, the images in $I$ are restored and we use $Restore$ to restore images in the white box. The results are shown in the right panel of Figure~\ref{realData2}. We can find that the spatial resolution of the observed images has been improved and the difference of image quality between the validation part and the training part is very small. The MFGS is increased from 0.78 to around 0.89 for this image.\\
\section{Conclusions} \label{sec:con}
As more and more high resolution solar images are obtained, we propose a pure data based image restoration method to make better use of these data. We assume texture features of solar images in the same wavelength are multi-fractals and use a deep neural network -- CycleGAN to restore blurred images, with several high resolution images from the same telescope as references. Our method does not need paired images as the training set. Instead, with only several high resolution images observed in the same wavelength, our method can give promising restoration results for every frame of real observation data without any additional instruments. We use simulated blurred images to test our algorithm. We compare the reconstructed images with real images and find that the MFGS has been increased with our method. Besides, we further use our algorithm to restore real observation images. Although the image quality increased by our method is slightly smaller than the speckle reconstructed images, our method can restore every frame of blurred images, while the speckle reconstructed method use a lot of blurred images (dozens or hundreds) and only obtains one frame of restored image. Our method is suitable for future observation data obtained by seeing-limited telescope or telescope with ground layer adaptive optic systems. Because our method does not have any prior assumption of the degradation process, it can also be used to restore images of other astronomical objects with features, such as galaxies, nebulae or supernova remnants.
\acknowledgments
The authors are grateful to the anonymous referee for his or her comments and suggestions, which have greatly improved the quality of this manuscript. The authors would like to thank Dr. Yongyuan Xiang and Professor Kaifan Ji from Yunnan Astronomical Observatory for their suggestions and providing solar observation data from the NVST. P.J. would like to thank Professor Hui Liu and Professor Zhong Liu from Yunnan Observatory, Professor Yong Zhang from Nanjing Institute of Astronomical Optics and Technology, Dr. Alastair Basden, Dr. Tim Morris, Dr. James Osborn and Dr. Matthew Townson from Durham University, Dr. Yang Guo and Dr. Qi Hao from Nanjing University and Dr. Qinmin Zhang from Purple Mountain Observatory who provide very helpful suggestions for this paper. This work is supported by National Natural Science Foundation of China (NSFC)(11503018), the Joint Research Fund in Astronomy (U1631133) under cooperative agreement between the NSFC and Chinese Academy of Sciences (CAS). PJ is supported by the China Scholarship Council to study at the University of Durham. The data used in this paper were obtained by the New Vacuum Solar Telescope in Fuxian Solar Observatory of Yunnan Astronomical Observatory, CAS.\\

\appendix
\section{Detail Structure of the CycleGAN} \label{app}
\startlongtable
\begin{deluxetable}{cccc}
\tablecaption{Structure of the Generator. Conv2d is standard convolutional layer. IN is a instance normalization layer which will normalize each image through $I=(I-\overline{I})/Var(I)$, where $\overline{I}$ and $Var(I)$ are mean value and variance of I. ReLu is the activation function. ResidualBlock is special structure of neural network and the input will feed into its output and the first layer as defined in Table~\ref{tab3}. ConvT2d is a transposed convolutional layer which will up sample the input data through learnable weights.\label{tab1}}
\tablehead{
\colhead{Type} & \colhead{kernel size $\backslash$ stride} & \colhead{output}
}
\startdata
Conv2d & $7\times 7 \backslash 1$ & $256\times 256\times 64$ \\
IN &    &  $256\times256\times 64$\\
ReLu&   & $256\times256\times64$\\
Conv2d & $3\times3 \backslash 2$ & $128\times128\times128$\\
IN &  &  $128\times128\times128$\\
ReLu &  & $128\times128\times128$ \\
Conv2d & $3\times3 \backslash 2$ &  $64\times64\times256$\\
ResidualBlock & &$64\times64\times256$\\
ResidualBlock & &$64\times64\times256$\\
ResidualBlock & &$64\times64\times256$\\
ResidualBlock & &$64\times64\times256$\\
ResidualBlock & &$64\times64\times256$\\
ConvT2d & $3\times3 \backslash 2$ & $128\times128\times128$\\
IN & & $128\times128\times128$\\
ReLu & &  $128\times128\times128$\\
ConvT2d & $3\times3 \backslash 2$ & $256\times256\times64$\\
IN & & $256\times256\times64$\\
ReLu & &  $256\times256\times64$\\
Conv2d & $7\times 7 \backslash 1$ & $256\times 256\times 1$\\
\enddata
\end{deluxetable}

\startlongtable
\begin{deluxetable}{ccccc}
\tablecaption{Structure of the Discriminator. LeakyReLu is a leakyReLu activation function, which has small slope for negative values, and in this paper we use the negative slope of 0.2. Sigmoid is the output layer of the discriminator and is used for classification of the input signals. All other layers have the same definitions as those in Table ~\ref{tab1}.\label{tab2}}
\tablehead{
\colhead{Type} & \colhead{kernel size $\backslash$ stride} & \colhead{output}&  \colhead{negative slope}
}
\startdata
Conv2d & $4\times 4 \backslash 2$ & $128\times 128\times 64$ & \\
LeakyReLU	 &    &  $256\times256\times 64$ & 0.2 \\
Conv2d & $4\times4 \backslash 2$ & $64\times64\times128$ &\\
IN&  & $64\times64\times128$ & \\
LeakyReLU	 &    &  $64\times64\times128$ & 0.2 \\
Conv2d & $4\times4 \backslash 1$ & $64\times64\times256$ &\\
IN&  & $64\times64\times256$ & \\
LeakyReLU	 &    &  $64\times64\times256$ & 0.2 \\
Conv2d & $4\times4 \backslash 1$ & $64\times64\times1$ &\\
Sigmoid & & &
\enddata
\end{deluxetable}

\startlongtable
\begin{deluxetable}{ccc}
\tablecaption{Structure of the ResidualBlock. The residualblock was firstly introduced by \citet{He2015}. All the layers have the same definitions as those in Table ~\ref{tab1}. The input will feed into the first layer and the output simultaneously ($ResidualBlockOUT=OUT+INPUT$, where $OUT$ is the output of the last $IN$ layer, $INPUT$ is the input of the ResidualBlock and $ResidualBlockOUT$ is the output of this ResidualBlock). \label{tab3}}
\tablehead{
\colhead{Type} & \colhead{kernel size $\backslash$ stride}
}
\startdata
Conv2d & $3\times 3 \backslash 1$  \\
IN &    \\
ReLu&   \\
Conv2d & $3\times3 \backslash 1$ \\
IN &  \\
\enddata
\end{deluxetable}

\bibliographystyle{aasjournal}
\bibliography{ref}

\end{document}